\documentclass[aps,preprint]{revtex4}%
\usepackage{amsfonts}
\usepackage{amsmath}
\usepackage{amssymb}
\usepackage{graphicx}%
\providecommand{\U}[1]{\protect\rule{.1in}{.1in}}

\begin{document}
\preprint{ }
\title[Short title for running header]{Comments on \textquotedblleft On the Consistency of the Solutions of the Space
Fractional Schr\"{o}dinger Equation\textquotedblright\ [J. Math. Phys.
\textbf{53}, 042105 (2012)]}
\author{Sel\c{c}uk \c{S}. Bayin}
\affiliation{Institute of Applied Mathematics, METU Ankara TURKEY 06800}

\begin{abstract}
Recently we have reanalyzed the consistency of the solutions of the space
fractional Schr\"{o}dinger equation found in a piecewise manner, and showed
that an exact and a proper treatment of the\ relevant integrals prove that
they are consistent. In this comment, for clarity, we present additional
information about the critical integrals and describe how their analytic
continuation is accomplished. [doi: 10.1063/1.4705268]

(Journal Ref.: JMP, \textbf{53}, 084101 (2012))

\end{abstract}
\maketitle

In a recent article [1], we have reanalyzed the consistency problem of the
solutions of the space fractional Schr\"{o}dinger equation found in a
piecewise fashion and showed that a proper treatment of the relevant
integrals, in contrast to the claims of Jeng et.al. [2], proves that they are
consistent. A crucial step of our proof was the analytic continuation of
certain integrals in evaluating their Cauchy principal values. In order to
remove any question marks about how analytic continuation is accomplished, we
present some additional details.

The space fractional Schr\"{o}dinger equation is written by using the Riesz
derivative, $R_{x}^{\alpha},$ which is defined with respect to the Fourier
transform
\begin{equation}
\mathcal{F}\left\{  R_{x}^{\alpha}\left(  g(x)\right)  \right\}
=-\frac{\left[  \left(  iq\right)  ^{\alpha}+\left(  -iq\right)  ^{\alpha
}\right]  }{2\cos\left(  \alpha\pi/2\right)  }G(q),\text{ }1<\alpha\leq2,
\end{equation}
where $\mathcal{F}\left\{  g(x)\right\}  =G(q).$ Taking the inverse, gives the
Riesz derivative as [3,4]%
\begin{equation}
R_{x}^{\alpha}\left(  g(x)\right)  =-\frac{1}{2\pi}\int_{-\infty}^{\infty
}\frac{\left[  \left(  iq\right)  ^{\alpha}+\left(  -iq\right)  ^{\alpha
}\right]  }{2\cos\left(  \alpha\pi/2\right)  }\ G(q)e^{iqx}dq.
\end{equation}
Since in equation (2) $q$ is real, using the relation%
\begin{equation}
\left[  \left(  iq\right)  ^{\alpha}+\left(  -iq\right)  ^{\alpha}\right]
=\left\vert q\right\vert ^{\alpha}2\cos\left(  \alpha\pi/2\right)  ,
\end{equation}
it is customary to write the Riesz derivative as%
\begin{equation}
R_{x}^{\alpha}\left(  g(t)\right)  =-\frac{1}{2\pi}\int_{-\infty}^{\infty
}\left\vert q\right\vert ^{\alpha}G(q)e^{i\omega x}dq.
\end{equation}
In this regard, Equation (12) of the paper [1] was written as%
\begin{equation}
\psi_{1}(x)=-\frac{AD_{\alpha}}{\pi E_{1}}\left(  \frac{\pi\hslash}%
{2a}\right)  ^{\alpha}\int_{-\infty}^{+\infty}dq\frac{\left\vert q\right\vert
^{\alpha}\cos\left(  \pi q/2\right)  }{\ q^{2}-1}e^{i\pi qx/2a},
\end{equation}
where $q$ is real.

To evaluate the Cauchy principal value of the above integral, we make use of
the Cauchy integral theorem, which requires that the integrand be analytic in
and on the closed contour described in [1]. Since $\left\vert q\right\vert
^{\alpha}$ is not analytic in the complex $q-$plane, the proper analytic
continuation of this integral is accomplished by going to the original form of
the Riesz derivative [Eq. (2)], and by writing Equation (5) with the
replacement
\begin{equation}
\left\vert q\right\vert ^{\alpha}\rightarrow\frac{\left[  \left(  iq\right)
^{\alpha}+\left(  -iq\right)  ^{\alpha}\right]  }{2\cos\left(  \alpha
\pi/2\right)  },
\end{equation}
and then performing analytic continuation. Similarly, analytic continuation of
Equation (25) of the paper is performed.

It is important to note that Equation (9) of [1]:%
\begin{equation}
\psi_{1}(x)=\left\{
\begin{tabular}
[c]{ccc}%
$A\cos\left(  \frac{\pi x}{2a}\right)  $ & if & $\left\vert x\right\vert <a$\\
&  & \\
$0$ & if & $\left\vert x\right\vert \geqslant a$%
\end{tabular}
\ \ \ \ \ \ \ \ \ \ \right.  .
\end{equation}
and equation (5), which are claimed to be inconsistent [2], are basically the
same wave function. Equation (5) is just the integral representation of the
wave function in Equation (7). We showed that when the integral is evaluated
properly, they indeed agree with each other [1]. This is true not just for the
ground state but for all the other states and also in general.

\end{document}